\documentclass[showpacs,aps,twocolumn,floatfix]{revtex4}

\usepackage{bm}
\usepackage{amsmath}
\usepackage{graphicx}
\usepackage{subfigure}
\usepackage[usenames,dvipsnames]{color}
\definecolor{darkblue}{RGB}{0,0,196}
\usepackage[colorlinks=true,linkcolor=darkblue,citecolor=darkblue,urlcolor=darkblue]{hyperref}

\usepackage{xcolor}

\usepackage{setspace}
\usepackage{footmisc}
\usepackage[makeroom]{cancel}
\usepackage{comment}
\usepackage{lineno}

\usepackage{graphicx}
\usepackage{amsmath,bbm}
\usepackage{amssymb,bm}
\usepackage{yfonts}

\usepackage{bm}
\usepackage{amsmath}
\usepackage{graphicx}
\usepackage{subfigure}
\usepackage[usenames,dvipsnames]{color}
\definecolor{darkblue}{RGB}{0,0,196}
\usepackage[colorlinks=true,linkcolor=darkblue,citecolor=darkblue,urlcolor=darkblue]{hyperref}
\usepackage{setspace}
\usepackage{hyperref}
\usepackage{footmisc}
\usepackage[makeroom]{cancel}
\usepackage{comment}
\usepackage{lineno}

\def\be{\begin{equation}}
\def\ee{\end{equation}}
\def\ba{\begin{eqnarray}}
\def\ea{\end{eqnarray}}

\usepackage{graphicx}
\usepackage{amsmath,bbm}
\usepackage{amssymb,bm}
\usepackage{yfonts}

\begin{document}
\title{Event multiplicity, transverse momentum and energy dependence of charged particle production, and system thermodynamics in $pp$ collisions at the Large Hadron Collider}
\author{Rutuparna Rath$^{a}$}
\author{Arvind Khuntia$^{b}$\footnote{e-mail: $Arvind.Khuntia@cern.ch$}}
\author{Raghunath Sahoo$^{a}$\footnote{Corresponding author: $Raghunath.Sahoo@cern.ch$}}
\author{Jean Cleymans$^{c}$\footnote{e-mails: $ jean.cleymans@gmail.com; jean.cleymans@uct.ac.za$}}

\affiliation{$^{a}$Discipline of Physics, School of Basic Sciences, Indian Institute of Technology Indore, Simrol, Indore 453552, India}
\affiliation{$^{b}$The H. Niewodniczanski Institute of Nuclear Physics,Polish Academy of Sciences, PL-31342 Krakow, Poland}
\affiliation{$^{c}$UCT - CERN Research Centre, Physics Department, University of Cape Town, Rondebosch - 7701, South Africa}

\begin{abstract}
 
 In the present work, we study the recent collision energy and multiplicity dependence 
 of the charged particle 
transverse momentum spectra as measured by the ALICE collaboration in $pp$ collisions at $\sqrt{s}$ = 5.02 and 13 TeV using 
the  non-extensive Tsallis distribution and the 
 Boltzmann-Gibbs Blast Wave (BGBW) model. A thermodynamically consistent 
form of the Tsallis distribution is used to extract the kinetic freeze-out parameters  from the transverse momentum spectra of
 charged particles at mid-rapidity. In addition, a comprehensive study of fitting 
 range dependence of transverse momentum spectra on the freeze-out parameters is done using Tsallis statistics. 
The applicability of BGBW model is verified by fitting the transverse momentum spectra   of the bulk part ($\sim 2.5~ {\rm GeV}/c$)
 for both 5.02 and 13 TeV energies and also in different multiplicity classes. The radial flow, $<\beta>$ is almost independent of
 collision energy and multiplicity whereas the behavior of kinetic freeze-out temperature significantly depends on multiplicity classes. 
It is found that the Tsallis distribution generally leads to a better description for the complete transverse momentum spectra  whereas the BGBW model explains the bulk part of the system.

\end{abstract}
 
\pacs{}
\date{\today}
\maketitle 

\section{Introduction}

Transverse momentum spectra of final state particles produced in high energy collisions provide important information about
 the dynamics of the produced systems. Previously  $pp$ collisions at the
 Relativistic Heavy-ion Collider (RHIC)
 and the Large Hadron Collider (LHC) provided a baseline for the understanding of particle production in heavy-ion collisions. However, recent results on final state multiplicity dependence of some of the observables in $pp$ collisions at $\sqrt{s} = 7$ TeV, which showed similar heavy-ion-like results at high multiplicities  \cite{ALICE:2017jyt}, have led to the question 
whether $pp$ at the LHC can be used as a baseline study to understand medium formation in heavy-ion collisions or one should look for QGP-droplets in high-multiplicity $pp$ collisions at the LHC energies \cite{Sahoo:2019ifs}. The multi-partonic
 interactions in high multiplicity $pp$ collisions could result in the formation of more number of final state hadrons, which might favor thermalisation in the system through transfer of momenta in multiple collisions at the partonic level. Thus such a complicated system can be explained by statistical models with few parameters.
 The study of transverse momentum spectra of final state particles provides information about the kinetic freeze-out processes in high energy 
collisions. 
The use  of  Tsallis non-extensive statistics incorporates a power-law distribution, which has a dominant contribution at higher
collision energies, making a better description of the transverse momentum spectra of charged particles and identified particles up to very 
 high  values of 
 $p_{\rm T}$ \cite{worku1,worku2,cleymans1,azmi-cleymans,wilk2,wilk3,wilk5,wilk6,Urmossy,Khuntia:2018znt,Khuntia:2017ite,tsallis-poland,tsallis1,Bhattacharyya:2015hya}.
 This description can be understood as containing
  temperature fluctuations in the Boltzmann-Gibbs sense either on an event-by-event basis or within the same event \cite{Bhattacharyya:2015nwa}. 
This fluctuation in the temperature is directly related to the non-extensive parameter $q$ \cite{Wilk:1999dr} and tells us about the departure of the 
system from an equilibrium state. To have a link between the phenomenological Tsallis distribution and the second law of thermodynamics, recently a $q$-dual entropy has been introduced \cite{Parvan:2019bga}. In this approach, the zeroth term approximation appears to be the presently used Tsallis non-extensive statistics. In the present analysis, we have explicitly checked the variation of the non-extensive parameter $q$ and the  Tsallis
 temperature with the fitting range involved for the transverse momentum spectra of charged particle produced in high energy $pp$ collisions at
 LHC energies \cite{Acharya:2019mzb}.

As an alternative, perhaps more standard description, we have used a blast-wave model based on
 collective flow in small systems. 
 One  way to extract  information about the radial flow is by studying the transverse momentum spectra of charged
 particles with the use of the BGBW model, which incorporates the radial flow into the Boltzmann-Gibbs distribution function. 
The BGBW model is quite good in explaining the bulk part of the system, however it fails at low-$p_{\rm T}$  which could possibly  be 
due to the decays of hadronic resonances. In this paper we extract the information about the kinetic freeze-out temperature and radial flow from
 the charged particle $p_{\rm T}$-distributions at the highest LHC energies as a function of multiplicity. We compare the Tsallis
 non-extensive statistics and the BGBW model in detail in the next section.

The paper is organized as follows. In Sec. \ref{method}, we briefly recall the details of the thermodynamically consistent
 Tsallis distribution function and also the  BGBW model used to describe the charged particle $p_{\rm T}$-spectra at mid-rapidity 
produced in $pp$ collisions at $\sqrt{s} = $ 5.02 and 13 TeV. The results obtained using the Tsallis non-extensive statistics and BGBW model
 are discussed as a function of charged particle multiplicity and collision energy in Sec. \ref{res:dis}. Finally, in Sec. \ref{sum:con}, we
 present the summary of our results.

\section{Methodology}
\label{method}

\subsection{Non-extensive Tsallis statistics}
\label{sec:tsallis}

The   Tsallis non-extensive  distribution function accounts for the power-law contribution at high-$p_{\rm T}$ and this could be understood
as 
empirically taking of QCD contributions. It is given by:
\begin{align}
f(E)  \equiv \exp_{q}\left(-\frac{E-\mu}{T}\right) \label{tsallis}, 
\end{align}
where $E = \sqrt{p^2+m^2}$, is the energy and $\mu$ is the chemical potential of the system and 
the $\exp_{q}(x)$ has the following form:
\begin{equation}
\label{expq}
\exp_{q}(x) \equiv
  \begin{cases}
    [1+(q-1)x]^{1/(q-1)}       & \quad \text{if }  x > 0\\
    [1+(1-q)x]^{1/(1-q)}       & \quad \text{if }  x \le 0\\
  \end{cases}
\end{equation}
in the  limit, $q \rightarrow$ 1, Eq. \eqref{expq} reduces to the standard exponential function:
\begin{eqnarray*}
\lim_{q \to 1} \exp_q(x) \rightarrow \exp(x).
\end{eqnarray*}

The expressions of the relevant thermodynamic quantities like entropy, $S$, energy density, $\epsilon$ (= $E/V$), pressure, $P$, and 
particle number, $N$, are given below. Note that an addition power of $q$ in Eq. \eqref{tsallis} is necessary to make 
Tsallis statistics thermodynamically consistent \cite{worku1,worku2,azmi-cleymans}.

\begin{gather}
\begin{split}
S &= - gV\int\frac{d^3p}{(2\pi)^3}\left[f^{q}{\rm ln}_{q}f - f\right],\\
\epsilon &= g\int\frac{d^3p}{(2\pi)^3}~E~f^{q},\\
P &= g\int\frac{d^3p}{(2\pi)^3}\frac{p^{2}}{3E}~f^{q},\\
N &= gV\int\frac{d^3p}{(2\pi)^3} f^{q} \label{eq7},
\end{split}
\end{gather}

\noindent with $g$ being the degeneracy factor and $V$ is the (Tsallis) freeze-out volume. 

The  yield can be written in the following form, as deduced from Eq. (\ref{eq7}):
\begin{equation}
E \frac{d^{3}N}{dp^3} = gVE \frac{1}{(2\pi)^3}\left[1 + (q-1)\frac{E-\mu}{T}\right]^{-\frac{q}{q-1}},
\end{equation}
and can be written in the form of rapidity, $y$, transverse mass, $m_{\rm T}$, and transverse momentum, $p_{\rm T}$:  
\begin{equation}
\begin{split}
\left.\frac{d^{2}N}{dp_{\rm{T}}dy}\right|_{y = 0} =& gV\frac{p_{\rm{T}}m_{\rm{T}}\cosh y}{(2\pi)^2}\\
&\left[1 + (q - 1) \frac{m_{\rm{T}}\cosh y-\mu}{T}\right]^{-\frac{q}{q-1}} \label{eq9}.
\end{split}
\end{equation}

At LHC energies, assuming the chemical potential, $\mu\simeq 0$ and at mid-rapidity i.e, $y$ = 0 this reduces to: 

\begin{equation}
\begin{split}
\left.\frac{d^{2}N}{dp_{\rm T}dy}\right|_{y = 0} =& gV\frac{p_{\rm{T}}m_{\rm{T}}}{(2\pi)^2}\\
&\left[1 + (q - 1) \frac{m_{\rm {T}}}{T}\right]^{-\frac{q}{q-1}} \label{eq10}.
\end{split}
\end{equation}

Furthermore, integration over the transverse momentum of Eq. (\ref{eq10})  leads to~\cite{ryb}:
\begin{eqnarray}
\left.\frac{dN}{dy}\right|_{y = 0} &=& \frac{gV}{(2\pi)^2} \int_0^\infty p_{\rm {T}}dp_{\rm{T}}m_{\rm {T}} \left[1 + (q - 1) \frac{m_{\rm {T}}}{T}\right]^{-\frac{q}{q-1}}\nonumber\\
&=& \frac{gVT}{(2\pi)^2} \left[\frac{(2 - q)m^{2} + 2mT + 2T^{2}}{(2 - q)(3 - 2q)}\right]\nonumber\\
&&       \left[1 + (q - 1) \frac{m}{T}\right]^{-\frac{1}{q-1}}. \label{eq11}
\end{eqnarray}
This makes it possible to eliminate the volume $V$ in favor of the yield at mid-rapidity $dN/dy$.
Now we can write the invariant yield for the charged particles with assumption that it is dominated by the
 production of $\pi$, K and $p$ and in this case we have considered the weight factor 0.8, 0.12 and 0.08 for $\pi$, K and $p$,  respectively,  following the experimental yields of identified particles \cite{Acharya:2018orn}. Further to test the sensitivity of these fractions in a tolerance limit of experimental uncertainties, we have varied the pion fraction and found that the obtained fitting parameters are not significantly different.

\begin{eqnarray}
\label{eq_final}
\left.\frac{d^{2}N_{ch}}{dp_{\rm T}dy}\right|_{y = 0} &=& A \sum_{i=\pi, K, p}  w_i\frac{m_{\rm T_i}}{T} p_{\rm T}\nonumber\\
         &&\left[\frac{(2 - q)m_i^{2} + 2m_iT + 2T^{2}}{(2 - q)(3 - 2q)}\right]^{-1}\nonumber\\
&&\left[1 + (q - 1) \frac{m_i}{T}\right]^{-\frac{1}{1-q}} \nonumber\\
&&\left[1 + (q - 1) \frac{m_{\rm T_i}}{T}\right]^{-\frac{q}{q-1}},
\end{eqnarray}

\noindent here $A$ is the charged particle yield at mid-rapidity, $q$ is the non-extensive parameter  and $T$ is the Tsallis temperature of the system. Again to deal with the change from rapidity to pseudo-rapidity we can use the following relation:

\begin{eqnarray}
\frac{dN}{dp_{\rm T}~d\eta} = \sqrt{1 - \frac{m^2}{m_{\rm T}^2 \cosh^2y}}
\frac{dN}{dy dp_{\rm T}} ,
\end{eqnarray}

which, at mid-rapidity, becomes:

\begin{equation}
\frac{dN}{dp_{\rm T}~d\eta} = \frac{p_{\rm T}}{m_{\rm T}}\frac{dN}{dp_{\rm T}~dy} ,
\end{equation}

hence the Eq. \eqref{eq_final} becomes
\begin{eqnarray}
\label{eq_final:2}
\left.\frac{d^{2}N_{ch}}{dp_{\rm T}d\eta}\right|_{y = 0} &=& A \sum_{i=\pi, K, p}  w_i\frac{p_{\rm T}^2}{T} \nonumber\\
         &&\left[\frac{(2 - q)m_i^{2} + 2m_iT + 2T^{2}}{(2 - q)(3 - 2q)}\right]^{-1}\nonumber\\
&&\left[1 + (q - 1) \frac{m_i}{T}\right]^{-\frac{1}{1-q}} \nonumber\\
&&\left[1 + (q - 1) \frac{m_{\rm T_i}}{T}\right]^{-\frac{q}{q-1}}.
\end{eqnarray}

\subsection{Boltzmann-Gibbs Blast Wave (BGBW) Model}
\label{sec:BGBW}
As an alternative description, we compare the fits using the Tsallis distribution with
 the BGBW model.
The expression for the invariant yield in the framework of BGBW, where particles decouple from the
 system at a temperature $T_{kin}$  is given as follows~\cite{Schnedermann:1993ws}:
 
\ba
\label{bgbw1}
E\frac{d^3N}{dp^3}=D \int d^3\sigma_\mu p^\mu exp(-\frac{p^\mu u_\mu}{T}),
\ea
$\mathrm{where}~ p^\mu~=~(m_T{\cosh}y,~p_T\cos\phi,~ p_T\sin\phi,~ m_T{\sinh}y), \nonumber$
is the particle four-momentum and the four-velocity $u^\mu=\cosh\rho~(\cosh\eta,~\tanh\rho~\cos\phi_r,~\tanh\rho~\sin~\phi_r,~\sinh~\eta) \nonumber$.

Again, the kinetic freeze-out surface is parametrised as, 
\ba
d^3\sigma_\mu~=~(\cosh\eta,~0,~0, -\sinh\eta)~\tau~r~dr~d\eta~d\phi_r,
\ea

\noindent where, $\eta$ is the space-time rapidity. Now assuming Bjorken correlation in rapidity i.e, $y = \eta$~\cite{Bjorken:1982qr}, 
\ba
\label{boltz_blast}
\left.\frac{d^2N}{dp_{\rm{T}}dy}\right|_{y=0} = D \int_0^{R_{0}} r\;dr\;K_1\Big(\frac{m_{\rm{T}}\;\cosh\rho}{T_{kin}}\Big)I_0\nonumber\\
\Big(\frac{p_{\rm{T}}\;\sinh\rho}{T_{kin}}\Big),
\ea
here  $D$ is the normalization constant and $m_{\rm T}=\sqrt{p_{\rm{T}}^2+m^2}$ is the transverse mass while  
$K_{1}$ and $I_0$ are the modified Bessel functions, 
 $\rho={\tanh}^{-1}\beta$, with $\beta=\displaystyle\beta_s\;\Big(r/R_0\Big)^n$, is the velocity of the transverse expansion \cite{Huovinen:2001cy,Schnedermann:1993ws, Tang:2011xq}. Again $\beta_s$ is the maximum surface velocity and the exponent $n$ describes the evolution 
of the flow velocity from any arbitrary $r$ to $R$ ($r$ $<R$), with $R$ being the maximum radius at the kinetic freeze-out surface.

Now, the average of the transverse velocity can be evaluated as: 
\ba
<\beta> =\frac{\int \beta_s\xi^n\xi\;d\xi}{\int \xi\;d\xi}=\Big(\frac{2}{2+n}\Big)\beta_s,
\ea
where $\xi =r/R_0$.

For the charged particle transverse momentum distribution in case of BGBW we have considered the linear flow profile ($n$ = 1) and also
 weight factors 0.8, 0.12 and 0.08 are considered for $\pi$, K and $p$ respectively, which is the same procedure as used for the Tsallis distribution.


 \begin{table*}[htbp]
\caption[p]{Number of mean charged particle multiplicity density corresponding to different multiplicity classes in $pp$ collisions at $\sqrt{s}=$ 5.02 TeV \cite{Acharya:2019mzb}.}
\label{table:mult_info-5}
\begin{tabular}{|c|c|c|c|c|c|c|c|c|c|c|c|}
\hline
\multicolumn{2}{|c|}{${\bf Class ~name}$}&V0M1&V0M2&V0M3&V0M4&V0M5&V0M6&V0M7&V0M8&V0M9&V0M10\\
\hline
\multicolumn{2}{|c|}{ $  \bf \big<{\frac{dN_{ch}}{d\eta} } \big>$} &19.2$\pm$0.9&15.1$\pm$0.7&12.4$\pm$0.5&10.7$\pm$0.5&9.47$\pm$0.47&8.04$\pm$0.42&6.56$\pm$0.37&5.39$\pm$0.32&4.05$\pm$0.27&2.27$\pm$0.27\\
\hline
\end{tabular}
 \end{table*}
 
\begin{table*}[htbp]
\caption[p]{Number of mean charged particle multiplicity density corresponding to different multiplicity classes in $pp$ collisions at $\sqrt{s}=$ 13 TeV \cite{Acharya:2019mzb}.}
\label{table:mult_info-13}
\begin{tabular}{|c|c|c|c|c|c|c|c|c|c|c|c|}
\hline
\multicolumn{2}{|c|}{${\bf Class ~name}$}&V0M1&V0M2&V0M3&V0M4&V0M5&V0M6&V0M7&V0M8&V0M9&V0M10\\
\hline
\multicolumn{2}{|c|}{ $  \bf \big<{\frac{dN_{ch}}{d\eta} } \big>$} &26.6$\pm$1.1&20.5$\pm$0.8&16.7$\pm$0.7&14.3$\pm$0.6&12.6$\pm$0.5&10.6$\pm$0.5&8.46$\pm$0.40&6.82$\pm$0.34&4.94$\pm$0.28&2.54$\pm$0.26\\
\hline
\end{tabular}
 \end{table*}

\section{Results and discussion}
\label{res:dis}

\subsection{Non-extensive statistics}
\begin{figure}[!ht]
\begin{center}
\includegraphics[width=80mm,scale=0.5]{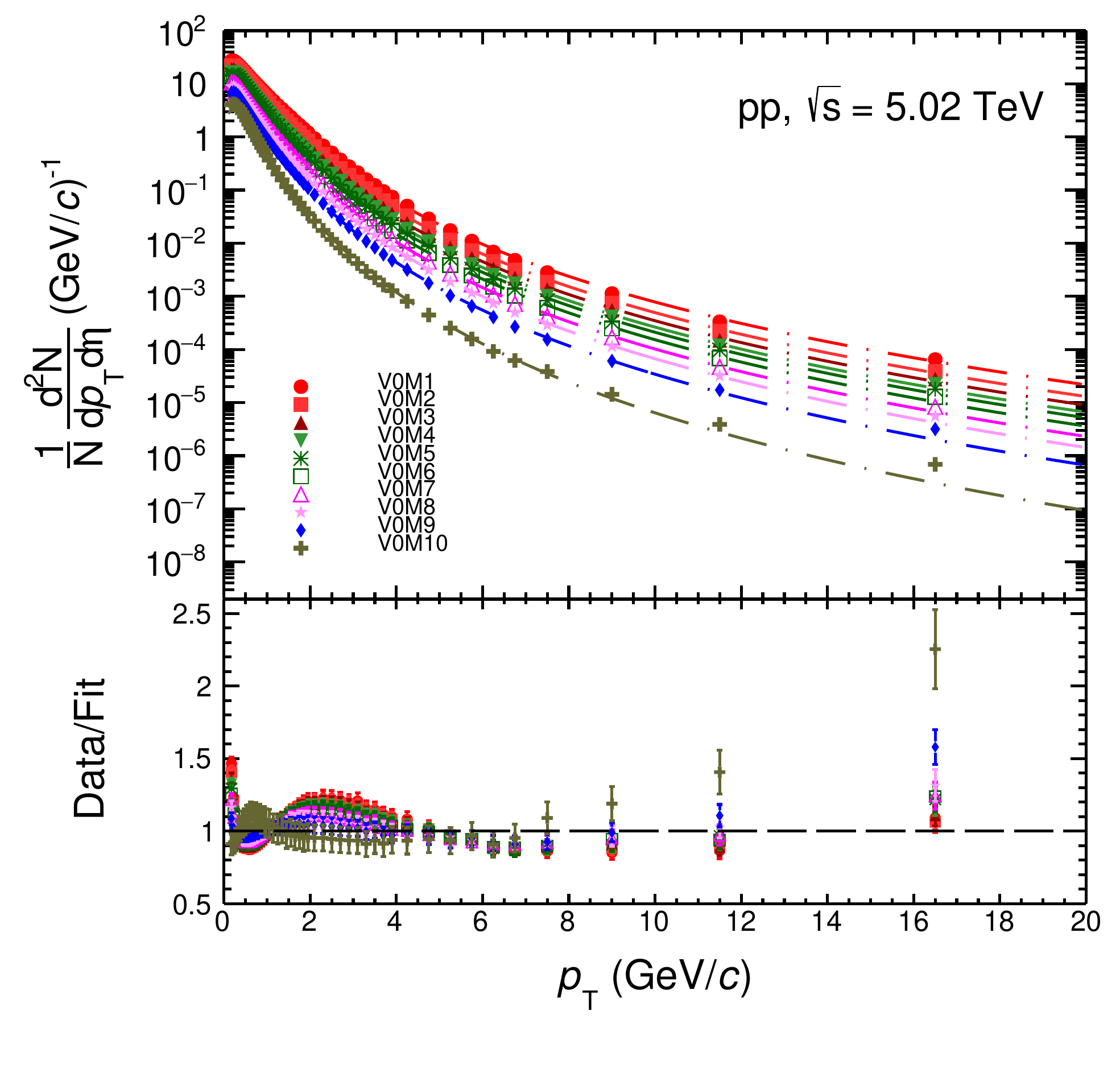}
\caption{(Color online) Charged particle spectra fit with Tsallis distribution function for $pp$ collisions at $\sqrt{s}$ = 5.02 TeV \cite{Acharya:2019mzb}.}
\label{fit:pp5:tsallis}
\end{center}
\end{figure}

\begin{figure}[!ht]
\begin{center}
\includegraphics[width=80mm,scale=0.5]{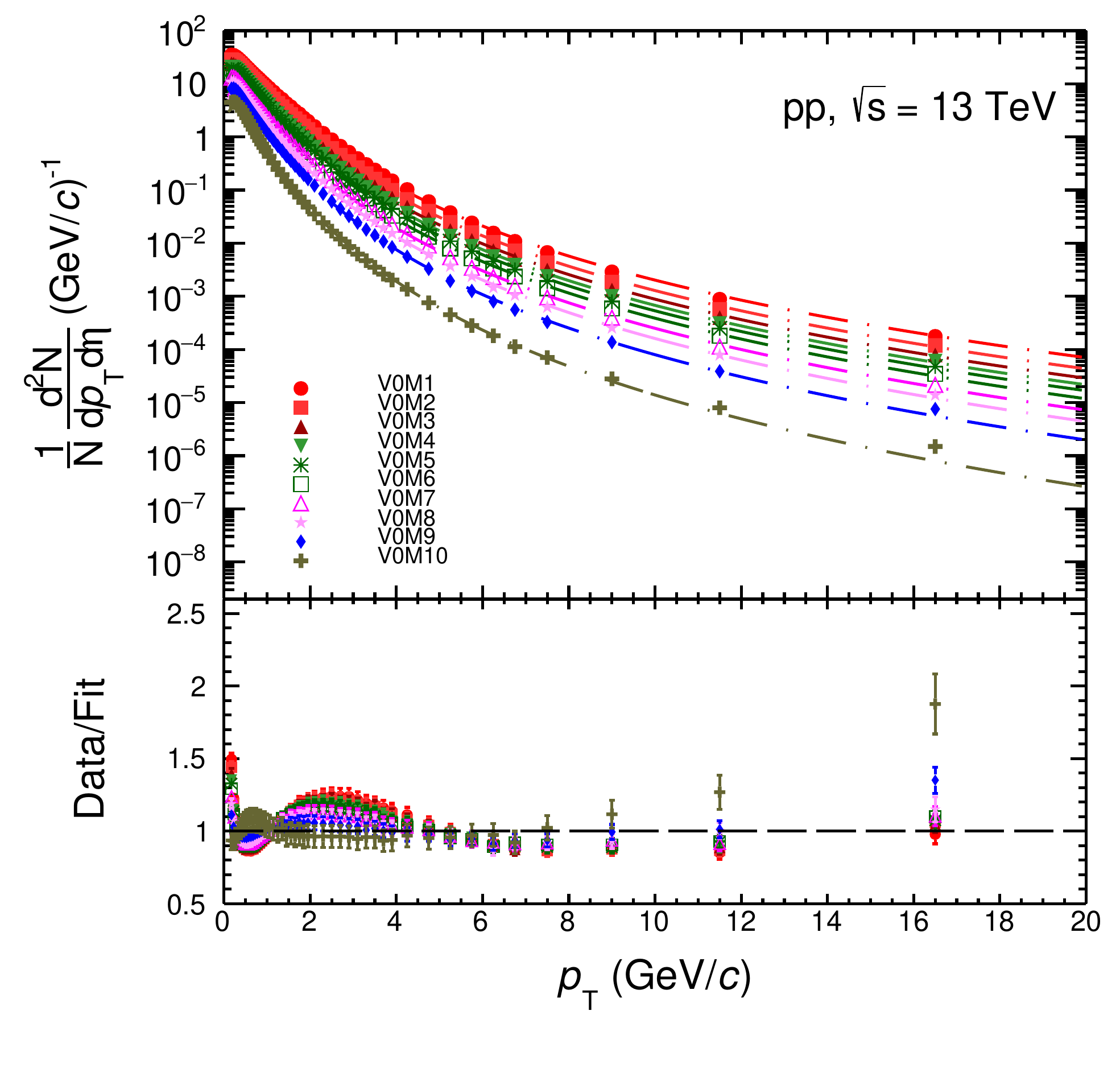}
\caption{(Color online) Charged particle spectra fit with Tsallis distribution function for $pp$ collisions at $\sqrt{s}$ = 13 TeV \cite{Acharya:2019mzb}.}
\label{fit:pp13:tsallis}
\end{center}
\end{figure}

The transverse momentum spectra ($p_{\rm T}$) of charged particles in $pp$ collisions at $\sqrt{s}$ = 5.02 and 13 TeV have been fitted with the 
 Tsallis non-extensive distribution upto 20 GeV/$c$ as shown in Fig. \ref{fit:pp5:tsallis} and \ref{fit:pp13:tsallis} \cite{Acharya:2019mzb}. The experimental data plotted here are with the systematic and statistical uncertainties added in quadrature. The fitting is also performed for different V0M event multiplicity classes measured at mid-rapidity ($|{\eta}| < $ 0.8) as given in Table \ref{table:mult_info-5} and \ref{table:mult_info-13}, at both  energies using  non-extensive statistics. The charged
 particle production might vary in different $p_{\rm T}$ regions and the $p_{\rm T}$-differential deviation of the fitting function from the 
charged particle transverse momentum spectra are shown in the bottom panel.

From the ratio between the experimental data points and the fit function, it is observed that the non-extensive statistics provides
 a good description of the charged particle transverse momentum spectra for the complete $p_{\rm T}$ region. However, in the  highest
transverse momentum region the fitting is better for the higher multiplicity classes as compared to the lower multiplicity classes. This behaviour gets reversed at the lower $p_{\rm T}$ region, which could be seen from the ratio plots.

\begin{figure}[!ht]
\begin{center}
\includegraphics[width=80mm,scale=0.5]{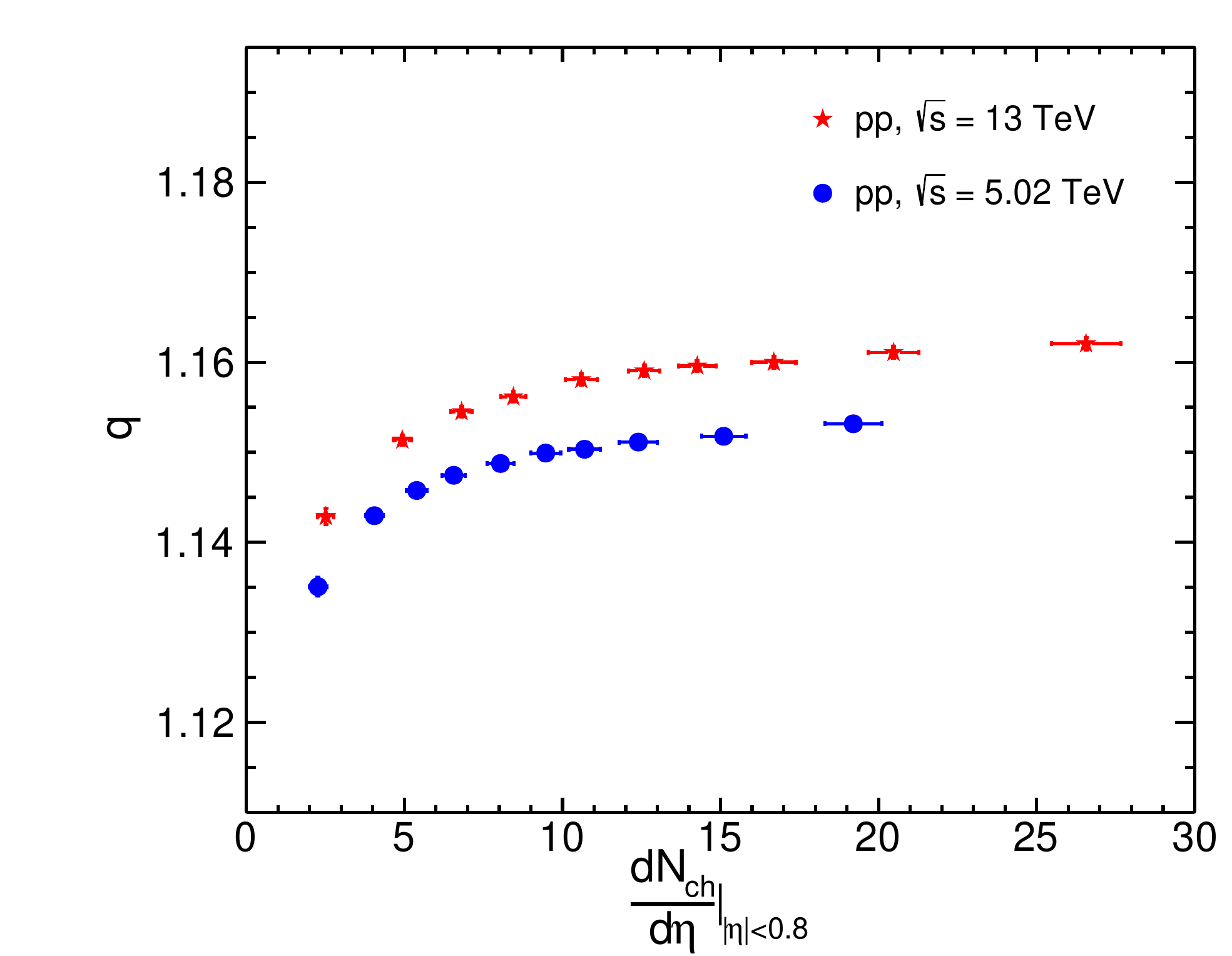}
\caption{ (Color online) Non-extensive parameter as a function of charged particle multiplicity and collision energy.}
\label{fit:q}
\end{center}
\end{figure}
The extracted non-extensive parameter, $q$, shown in Fig. \ref{fit:q} increases with multiplicity classes and
 remains almost constant above $\sim$$|dN_{ch}/d\eta|_{|\eta|<0.8} > $ 15. Such behaviour  has also been observed  for pions as a function of charged particle multiplicity in ref. \cite{Khuntia:2018znt}. The non-extensive parameter is higher for the 13 TeV and this can be understood as the contributions from the hard scatterings (jet contribution) in $pp$ collisions at $\sqrt{s} = $ 13 TeV is larger than at 5.02 TeV. Although MPIs contribute to high final state multiplicity,
 from the multiplicity dependence of $q$-parameter, one can conclude that in a jet-fragmentation rich environment, high-$p_{\rm T}$ particles come out of the system without much interaction. This could result in higher values of $q$ for high-multiplicity events.

\begin{figure}[!h]
\begin{center}
\includegraphics[width=80mm,scale=0.5]{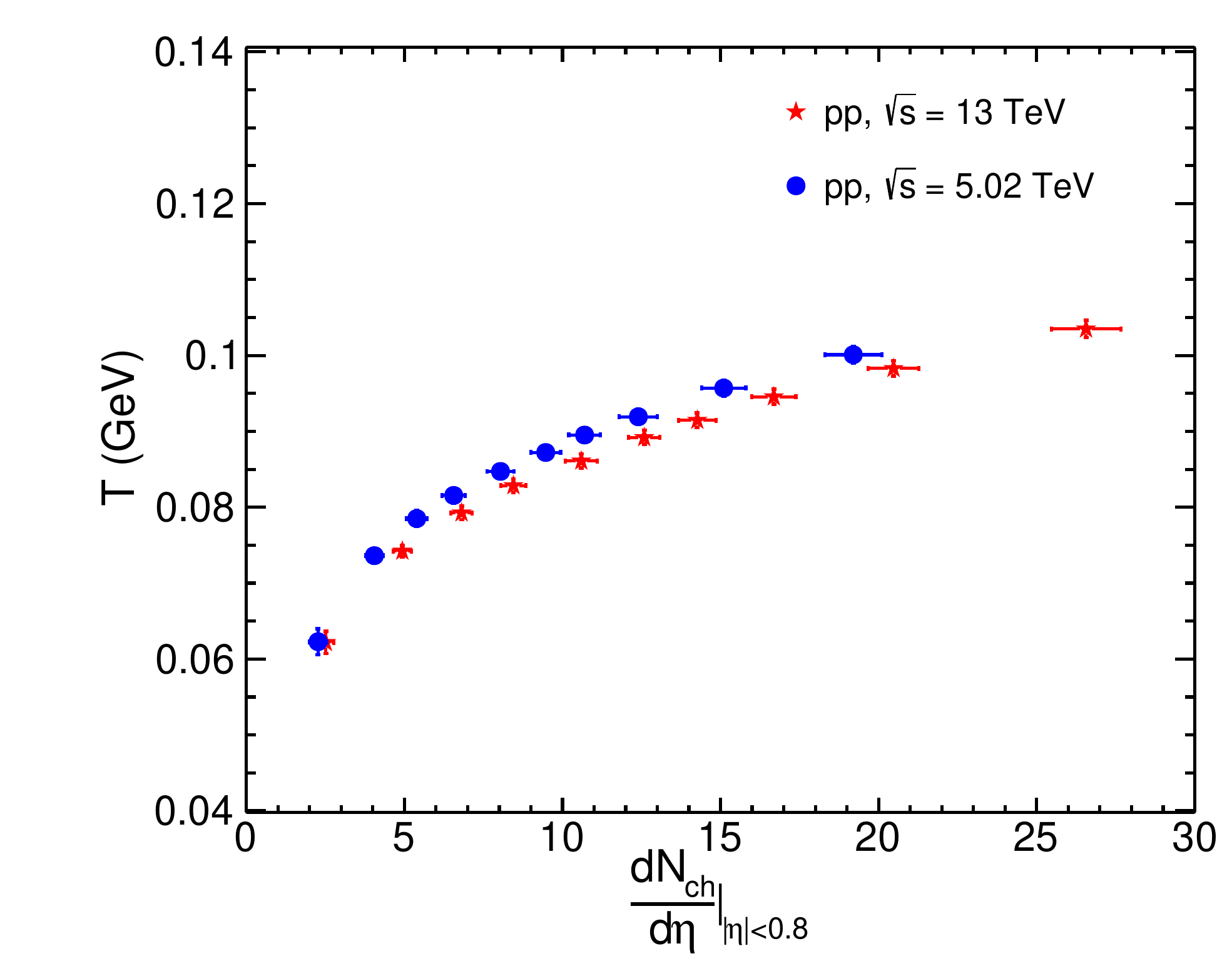}
\caption{ (Color online) Tsallis temperature as a function of charged particle multiplicity and collision energy.}
\label{fit:T}
\end{center}
\end{figure}

Furthermore, similar behavior is observed for the Tsallis temperature, $T$ and this shows a significant dependence on the multiplicity 
class as shown in Fig. \ref{fit:T}. 

\begin{figure}[!ht]
\begin{center}
\includegraphics[width=80mm,scale=0.5]{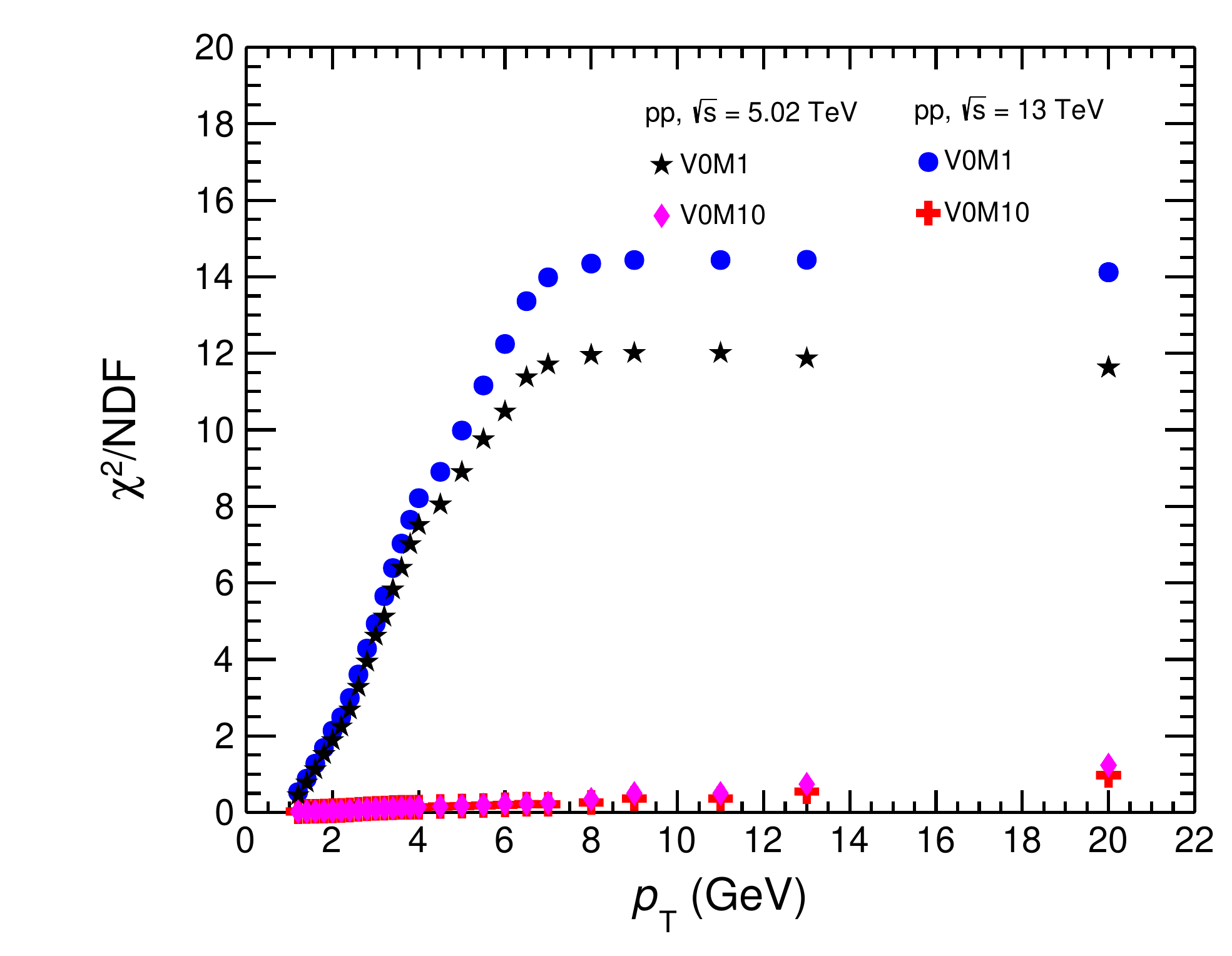}
\caption{ (Color online) $\chi^{2}/ndf$ obtained by varying the fitting ranges in the $p_{\rm T}$-spectra for $pp$ at $\sqrt{s}$ = 5.02 and 13 TeV for V0M1 and V0M10 classes.}
\label{fit:chi_diffpt}
\end{center}
\end{figure}

\begin{figure}[!ht]
\begin{center}
\includegraphics[width=80mm,scale=0.5]{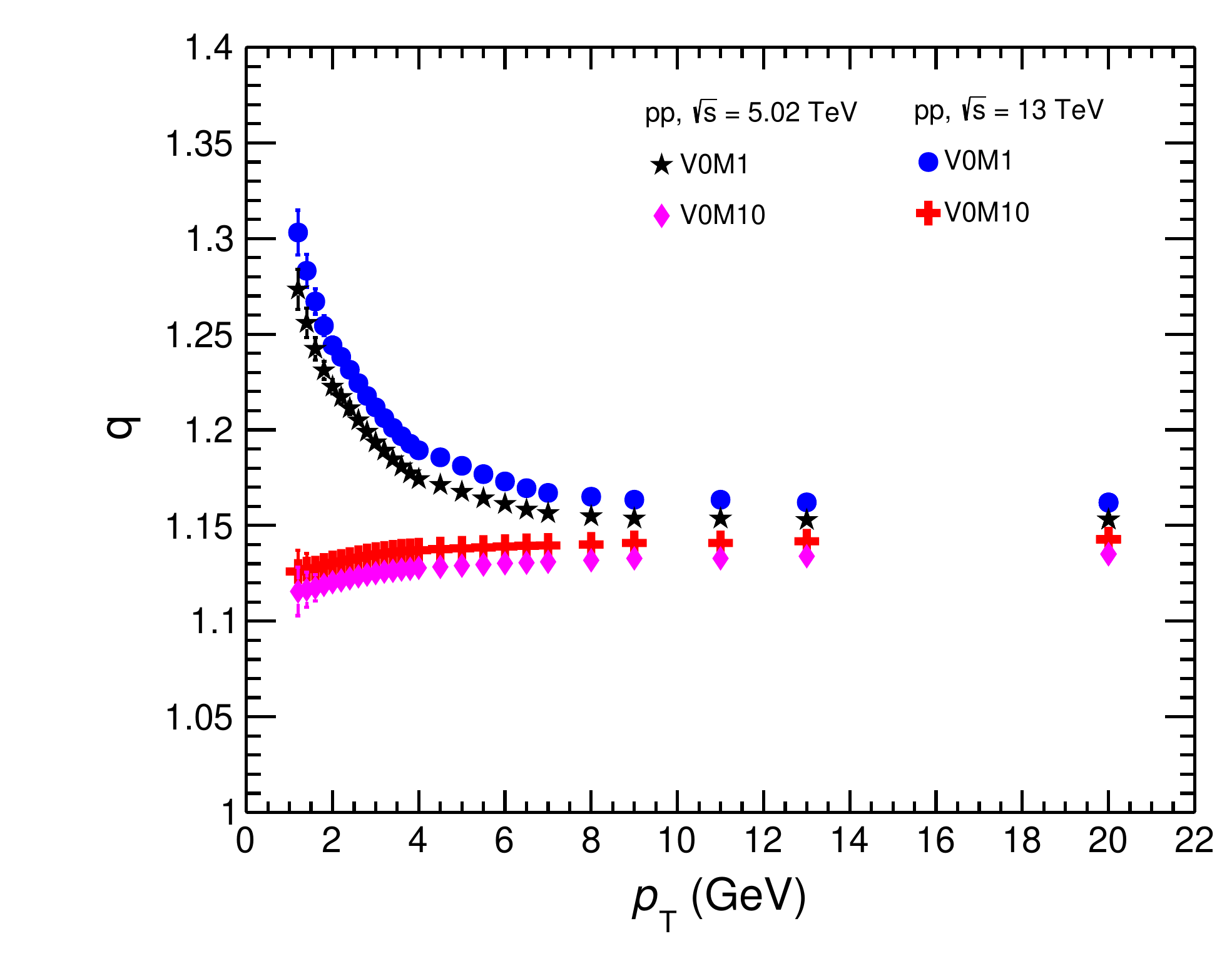}
\caption{ (Color online) Non-extensive parameter obtained by varying the fitting ranges in the $p_{\rm T}$-spectra for $pp$ at $\sqrt{s}$ = 5.02 and 13 TeV for V0M1 and V0M10 classes.}
\label{fit:q_diffpt}
\end{center}
\end{figure}

\begin{figure}[!ht]
\begin{center}
\includegraphics[width=80mm,scale=0.5]{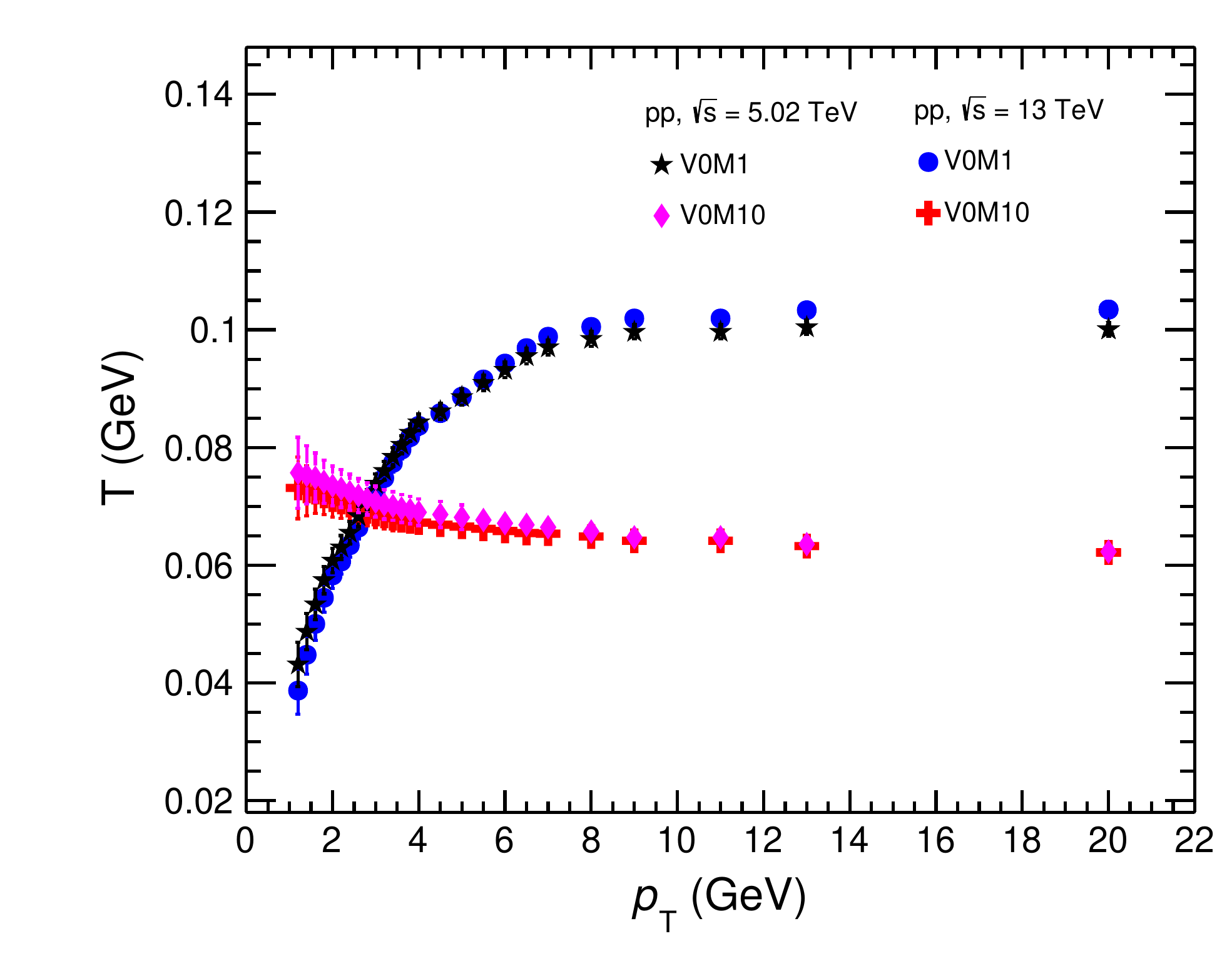}
\caption{ (Color online) Tsallis temperature parameter obtained by varying the fitting ranges in the $p_{\rm T}$-spectra for $pp$ at $\sqrt{s}$ = 5.02 and 13 TeV for V0M1 and V0M10 classes.}
\label{fit:T_diffpt}
\end{center}
\end{figure}

A comprehensive study of the extracted parameter dependence on the fitting range of charged particle transverse momentum spectra is
 performed in this work.  Here, we have considered the highest and the lowest charged particle multiplicity classes for both the energies. 
 The maximum value of transverse momentum in the fitting ranges from 1.2 to 20 GeV/$c$ and the extracted parameters are drawn and filled at the higher edge of the $p_{\rm T}$ as shown in Fig. \ref{fit:chi_diffpt}, \ref{fit:q_diffpt} and \ref{fit:T_diffpt}. The goodness of the fit i.e, $\chi^{2}/ndf$ as a
 function of fitting range is shown in Fig. \ref{fit:chi_diffpt}. In case of the lowest multiplicity class, for both the collision energies discussed here, the Tsallis distribution function seems to do a good job in describing the charged particle spectra. On the other hand, for the highest multiplicity class, as shown in the figure, for $p_{\rm T} \geq$ 2 GeV/c, the Tsallis description of the spectra becomes worsen. Although it is seen that the multiplicties achieved in heavy-ion collisions, Tsallis distribution function completely fails to describe the data, a smooth behaviour in multiplicity from $pp$ to heavy-ion collisions 
 is not envisaged, as the system dynamics becomes completely different. We have seen the onset of collectivity in high-multiplicity $pp$ collisions at the LHC energies \cite{ALICE:2017jyt}. Tsallis distribution doesn't account for this, which could be the reason of the above observation.

The  non-extensive parameters $q$ have been extracted by varying the fitting ranges and are shown in Fig. \ref{fit:q_diffpt}, for both
 V0M1 (high multiplicity) and V0M10 (low multiplicity) multiplicity classes.
The non-extensive parameter, $q$ shows a reverse trend for both the highest and the lowest multiplicity classes as a function of fitting range.

This is evident from the fact that the $p_{\rm T}$ spectra of the charged particles evolve with multiplicity class and the differential
 ratios of transverse momentum spectra with respect to minimum bias shows hardening of $p_T$-spectra with multiplicity
 classes \cite{Acharya:2019mzb}.  For the V0M1 class, the addition of high-$p_{\rm T}$ charged particles increases the 
 power-law contributions, which correspond to hard pQCD processes. However, this leads to the lowering of the values of $q$ with extending the fitting range in the non-extensive statistics for the highest multiplicity class. This may result from the bad $\chi^2/ndf$
 values for the V0M1 multiplicity class as a function of fitting range in $p_{\rm T}$. Contrary to the highest multiplicity class (V0M1), for the V0M10 class (the lowest multiplicity class), extending the fitting range has less power-law contribution compared to V0M1. This could be seen from the weak increase of  $q$-parameter as a function of fitting range in $p_{\rm T}$. For V0M10, $q$ increases with the fitting range in the non-extensive statistics approaching a saturation behavior beyond $p_{\rm T} \sim$ 8 GeV/c. The non-extensive parameter, $q$ doesn't depend on the fitting range above 8 GeV/$c$ for both V0M1 and V0M10 classes.

Similarly the Tsallis temperature, $T$, evolves with fitting range and has opposite behavior for V0M1 and V0M10 classes as
 shown in Fig. \ref{fit:T_diffpt}. The Tsallis temperature, $T$, increases with fitting range for V0M1 class, whereas the trend for 
V0M10 class is reverse and it decreases with the fitting range. Furthermore,  $T$ has a mild dependence on the fitting
 range for both 5.02 and 13 TeV. As it appears, $p_{\rm T} \sim$ 8 GeV/c seems to be a threshold for a different behavior in particle production.



\subsection{BGBW model}

\begin{figure}[!h]
\begin{center}
\includegraphics[width=80mm,scale=0.5]{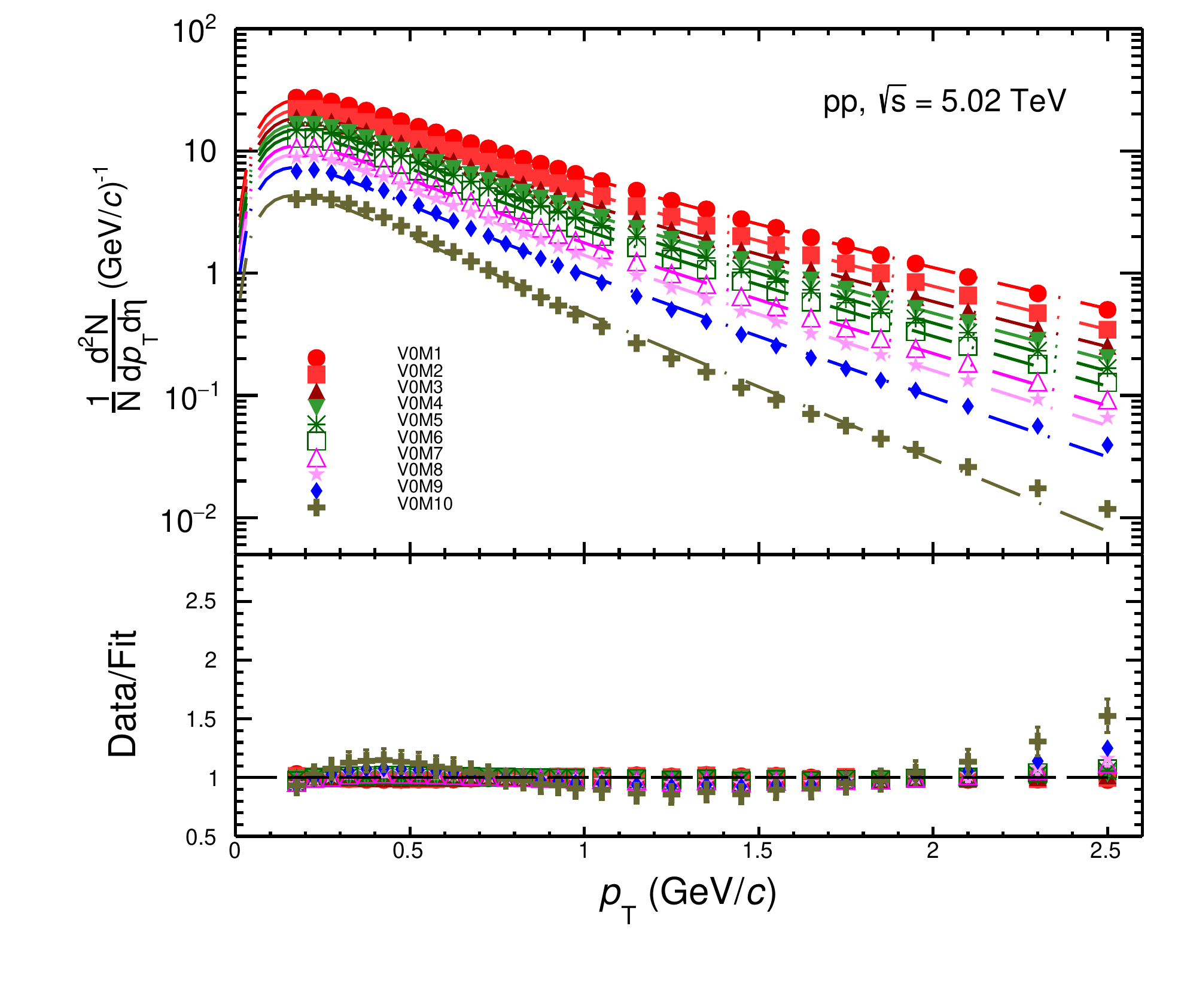}
\caption{(Color online) Charged particle spectra fit with BGBW function for $pp$ collisions at $\sqrt{s}$ = 5.02 TeV \cite{Acharya:2019mzb}.}
\label{fit:pp5:BGBW}
\end{center}
\end{figure}

\begin{figure}[!ht]
\begin{center}
\includegraphics[width=80mm,scale=0.5]{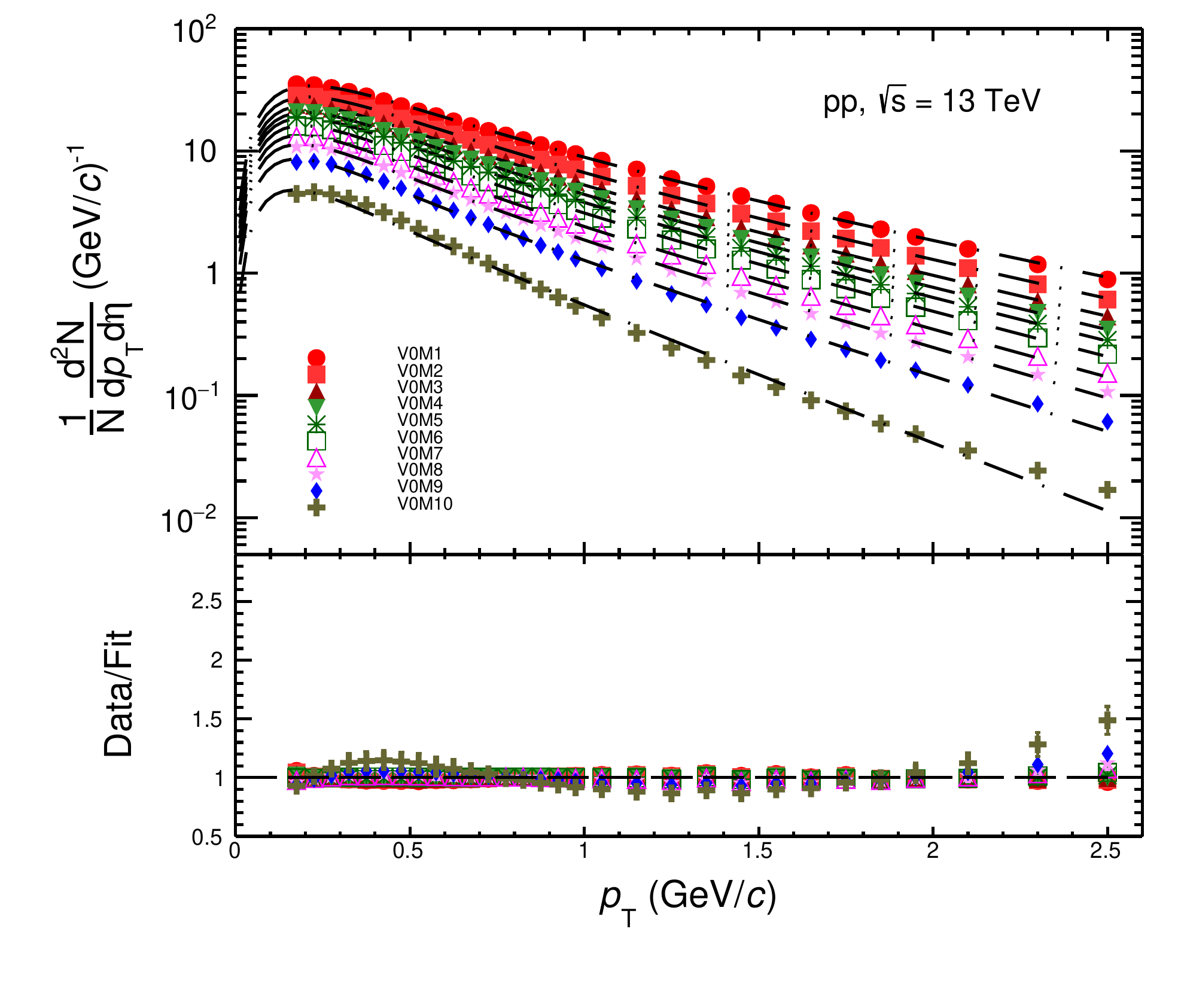}
\caption{(Color online) Charged particle spectra fit with BGBW function for $pp$ collisions at $\sqrt{s}$ = 13 TeV \cite{Acharya:2019mzb}.}
\label{fit:pp13:BGBW}
\end{center}
\end{figure}

\begin{figure}[!ht]
\begin{center}
\includegraphics[width=80mm,scale=0.5]{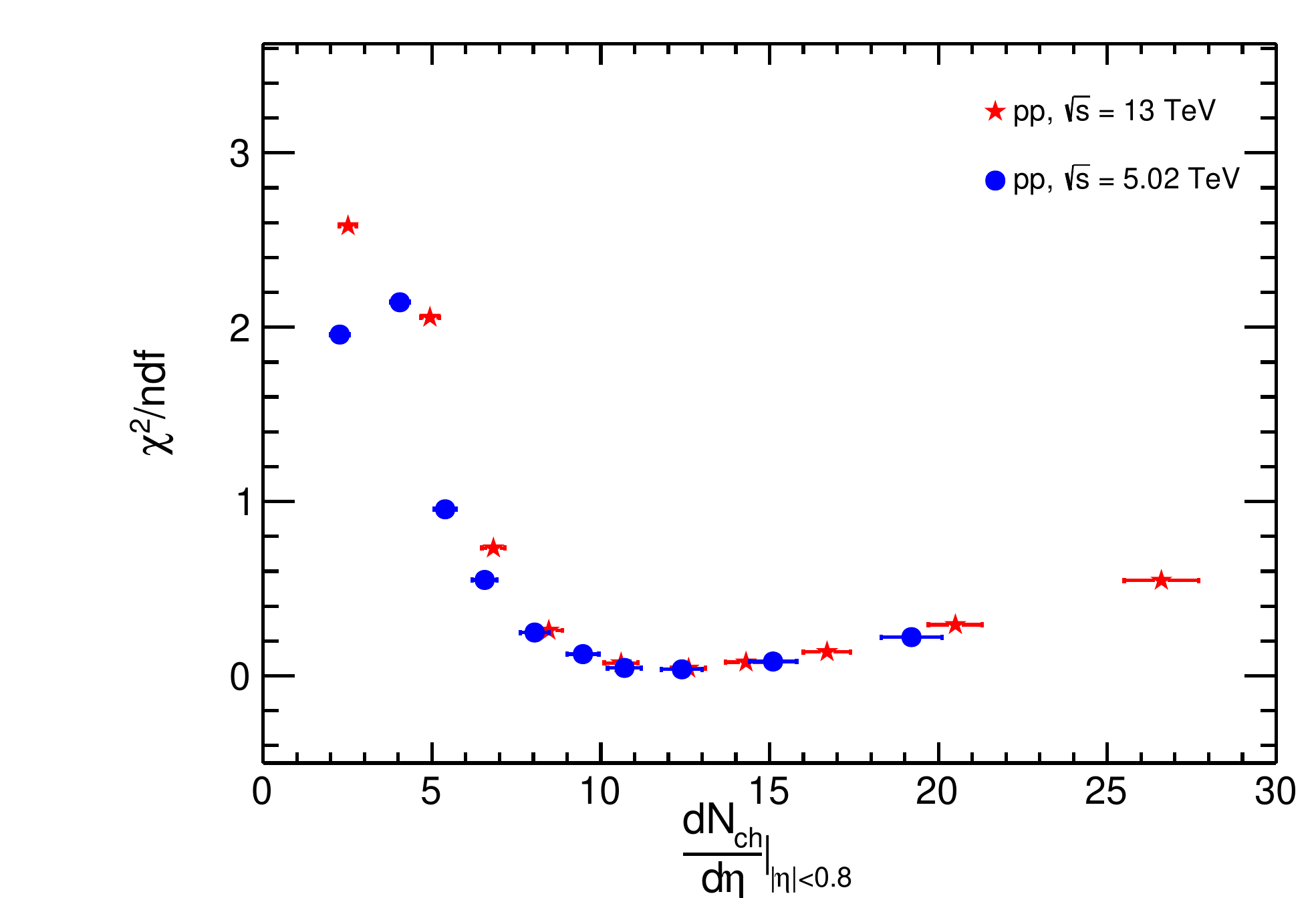}
\caption{ (Color online) The values of $\chi^2$ divided by the number of degrees of freedom
  as a function of the  charged particle multiplicity and collision energy for the BGBW analysis.}
\label{fig:BGBW:chi2}
\end{center}
\end{figure}

In this section, we analyse the bulk part (up to $\simeq$ 2.5 GeV/$c$) of the transverse momentum spectra of the charged particle
 using a BGBW model and characterize the system at the kinetic freeze-out with transverse collective flow velocity ($\beta$) and the
 kinetic freeze-out temperature ($T_{\rm kin}$). Fig. \ref{fit:pp5:BGBW} and \ref{fit:pp13:BGBW}, show the fit to the charged particle
 transverse momentum spectra in $pp$ collisions at $\sqrt{s}$ = 5.02 and 13 TeV for  different multiplicity classes.  The bottom panel 
shows the ratios between  the experimental data points and the BGBW function and the goodness of fit i,e. $\chi^2/ndf$ is shown
 by Fig. \ref{fig:BGBW:chi2}.

The BGBW model clearly  underestimates the charged particle transverse momentum spectra at low-$p_{\rm T}$ and thus the 
deviation of the BGBW function from the experimental data points at low-$p_{\rm T}$ arises which is possibly  due to the contributions from
 resonance decays. The BGBW fit is better for the higher charged particle multiplicity classes as compared to the lower multiplicity classes.

\begin{figure}[!ht]
\begin{center}
\includegraphics[width=80mm,scale=0.5]{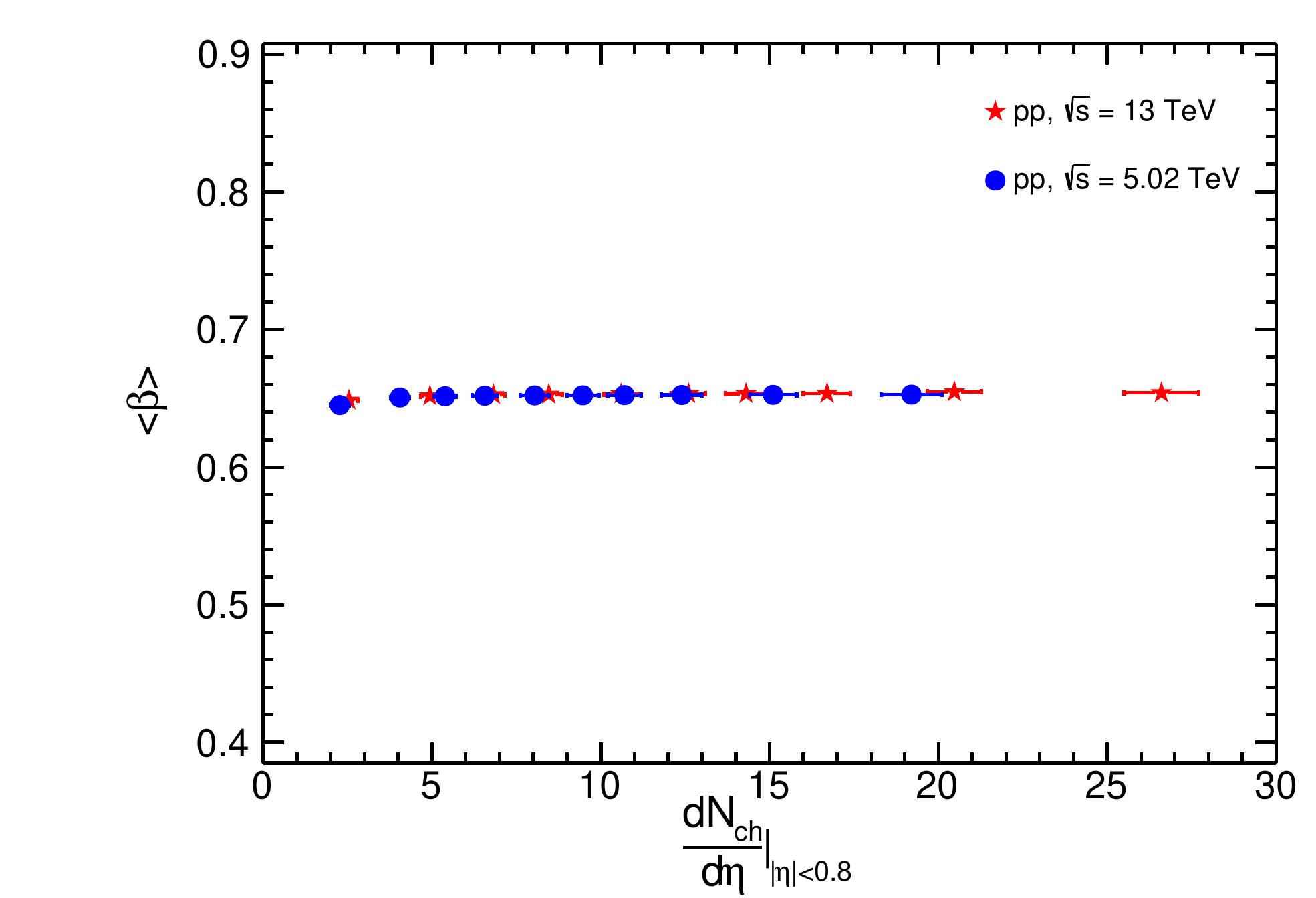}
\caption{ (Color online) Radial flow velocity parameter as a function of charged particle multiplicity and collision energy.}
\label{fig:BGBW:beta}
\end{center}
\end{figure}

\begin{figure}[!ht]
\begin{center}
\includegraphics[width=80mm,scale=0.5]{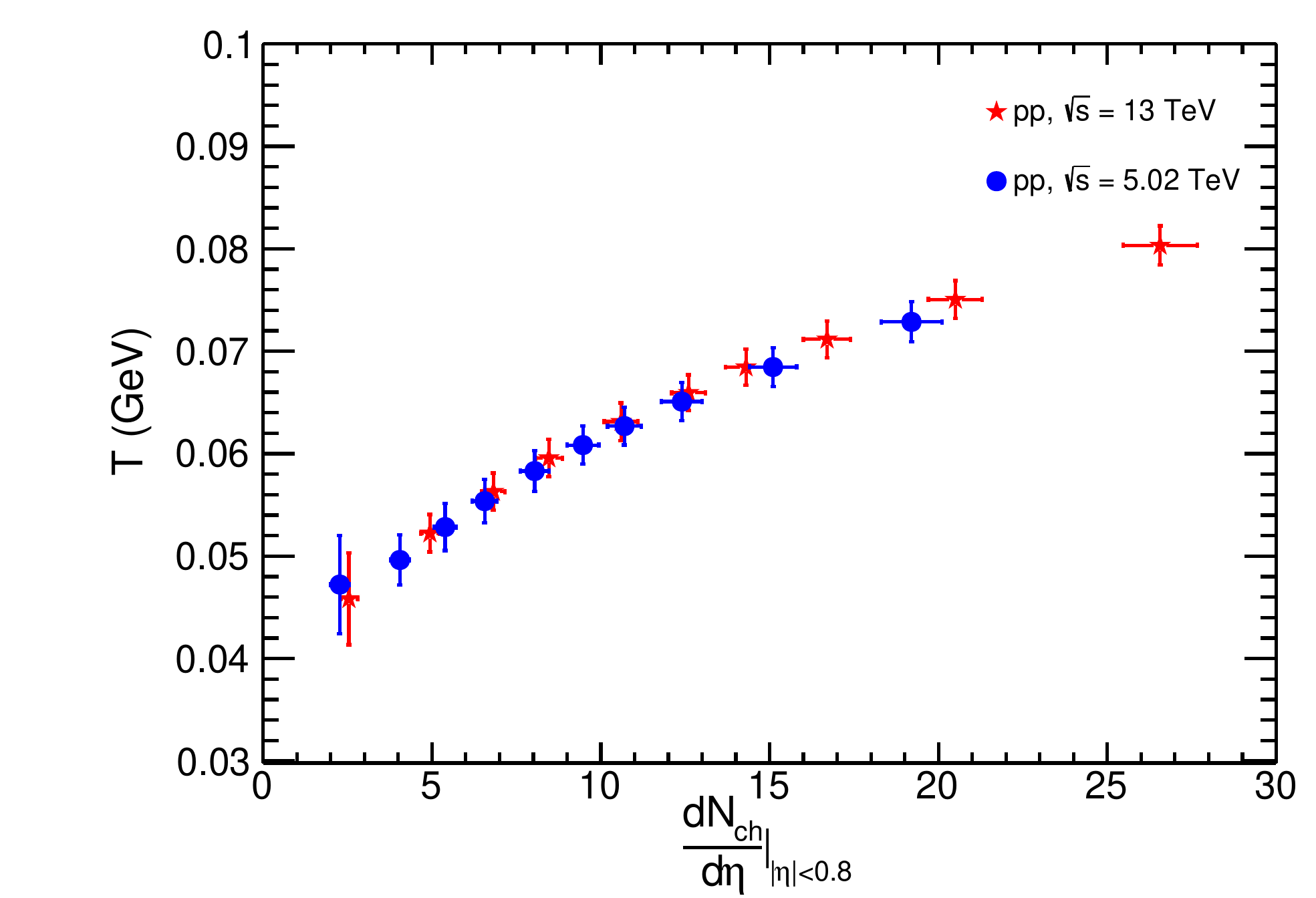}
\caption{ (Color online) Temperature as a function of charged particle multiplicity and collision energy.}
\label{fig:BGBW:T}
\end{center}
\end{figure}

The kinetic freeze-out parameters, the radial flow $(\beta)$ and kinetic freeze-out temperature $T_{\rm kin}$ are 
extracted using the BGBW model for various multiplicity classes and are shown in Fig. \ref{fig:BGBW:beta} and \ref{fig:BGBW:T}.
 The radial flow $(\beta)$ is almost independent of the collision energy and multiplicity classes. The kinetic freeze-out temperature, $T_{\rm kin}$ however shows a
 clear dependence on the multiplicity class. Furthermore, $T_{\rm kin}$ increases with charged particle multiplicity classes for both the collision energies. This trend of kinetic temperature can be understood from the hardening of charged particle spectra with event multiplicity classes. This is
 also reflected in the increase of $\langle p_{\rm T}\rangle$ with charged particle multiplicity \cite{Acharya:2019mzb}. 


\section{Summary}
\label{sum:con}

In the present work,  we have analyzed the charged particle transverse momentum spectra in $pp$ collisions at $\sqrt{s}=$ 5.02 and 13 TeV
 energies using the thermodynamically consistent Tsallis non-extensive statistics and, as a comparison, the  BGBW model. 
In both the cases, a weight factor
 of 0.8, 0.12 and 0.08 have been considered for $\pi$, K and $p$ respectively due to the fact that charged particles are mostly dominated
 by these three particles. A consistent picture of the extracted parameters emerges from the comparison to the experimental data as a
 function of charged particle multiplicity and collision energy and the important results are summarized below: 

\begin{itemize}
\item The non-extensive Tsallis distributions leads to a very good description of experimental data for the complete 
transverse momentum range in various multiplicity classes and collision energies. However, in the low transverse momentum region the fitting is better for the lower multiplicity classes as compared to the higher event multiplicity classes.

\item The deviation of the non-extensive parameter, $q$  from unity tells about the departure of the system from 
thermodynamic equilibrium and it increases with multiplicity classes and remains almost constant above $\sim$$|dN_{ch}/d\eta|_{|\eta|<0.8} > $ 15. Although this observation seems counter intuitive, it could be because of jet-fragmention contributing to the particle production making produced high-$p_{\rm T}$ particles come out of the system without interaction. An event shape dependence study on experimental data would be useful to have a better understanding of the underlying dynamics.

\item In line with other observations like multipartonic interactions dominantly contributing to the final state particle production \cite{Thakur:2017kpv}, and the observation of a thermodynamic limit for statistical ensembles to converge \cite{Sharma:2018jqf}, final state charged particle multiplicity $\sim$$|dN_{ch}/d\eta|_{|\eta|<0.8} > $ 15 appears to be a threshold number in particle production in hadronic collisions at the TeV energies.

\item Higher values of the non-extensive parameter is observed for 13 TeV as compared to 5.02 TeV and this might be understood as being due to the contributions from the hard scatterings in $pp$ at $\sqrt{s} = $13 TeV being higher as compared to 5.02 TeV.

\item Similar multiplicity dependent behavior for the Tsallis temperature, $T$ has been observed. We do observe a weak energy dependence of the Tsallis temperature, $T$ as a function of charged particle multiplicity. A correlation between $T$ and $q$ is clearly observed as a function of charged particle multiplicity.

\item Comprehensive study of the extracted parameters dependence on the fitting range of charged particle transverse momentum spectra is performed for the highest and the lowest charged particle multiplicity classes for both the energies.

\item The non-extensive parameter, $q$ shows a reverse trend for both the highest and the lowest multiplicity classes as a function of fitting range.

\item For the V0M1 class (corresponds to the highest multiplicity), the addition of  the high-$p_{\rm T}$ charged particles increases the power-law contributions. However, the non-extensive parameter, $q$ decreases with the fitting range in $p_{\rm T}$. This might result
from the fact that the Tsallis non-extensive statistics fails to describe the charged particle spectra for the V0M1 multiplicity class, as seen from the $\chi^2/ndf$ values. On the other hand, for the V0M10 class (corresponds to the lowest multiplicity), extending the 
fitting range has less power-law contribution (as compared to V0M1). The $q$ values for V0M10 show an increase with the fitting range in the non-extensive statistics.

\item $p_{\rm T} \sim$ 8 GeV/$c$, which shows a saturation behavior in $T$, $q$ and $\chi^2$/ndf seems to be a threshold for a different behavior in particle production. This needs a closure look.

\item Similarly, Tsallis temperature, $T$ evolves with fitting range and has opposite behavior for both V0M1 and V0M10 classes. The Tsallis
 temperature, $T$ increases with fitting range for V0M1 class, whereas the trend for V0M10 class is reverse with the fitting range. 

\item Furthermore, the bulk part (up to $\simeq$ 2.5 GeV/$c$) of the transverse momentum spectra of the charged particle has been
 characterized using BGBW model and extracted the transverse collective flow velocity ($<\beta>$) and the kinetic freeze-out 
temperature ($T_{\rm kin}$) as a function of charged particle multiplicity.

\item The collective radial flow velocity, $<\beta>$ is almost independent of the collision energy and multiplicity classes. The kinetic freeze-out temperature, $T_{\rm kin}$ however, shows a clear dependence on the multiplicity classes.

\end{itemize}

Conclusively, the BGBW explains the bulk part of the transverse momentum spectra and the description is better for the higher
 multiplicity classes, whereas the non-extensive Tsallis statistics describes the charged particle transverse momentum for the
 complete $p_{\rm T}$-range for the low multiplicity classes. In a jet rich environment, which might happen in high-multiplicity classes, the produced (mini)jets hardly interact with the system. Along with collectivity seen in these events, the $p_{\rm T}$
 spectra seem to show a deviation from the tendency of equilibration and hence $q$-values deviating from Boltzmann expectations.
 An event shape analysis on experimental data would be useful to understand the underlying dynamics, in view of a counter intuitive
 observation of high-multiplicity environment showing a away from equilibrium feature.
 
\section*{Acknowledgment}

The authors acknowledge the financial supports from ALICE Project No. SR/MF/PS-01/2014-IITI(G) of Department of Science $\&$ Technology, Government of India. RR acknowledges the financial support by DST-INSPIRE program of Government of India. Fruitful discussions with Dhananjaya Thakur is highly appreciated.

 \end{document}